\documentstyle[12pt]{article}
\begin{document}
\renewcommand{\theequation}{\thesection.\arabic{equation}}
\begin{titlepage}
\title{Maximal Acceleration Corrections to the \\
       Lamb Shift of Muonic Hydrogen }
\author{C.X. Chen, G. Papini\thanks{E-mail:
papini@cas.uregina.ca}, N. Mobed\thanks{E-mail: mobed@meena.cc.uregina.ca}, \\
{\em Department of Physics, University of Regina} \\
{\em Regina, Sask. S4S 0A2, Canada,}  \\
and \\
G. Lambiase\thanks{E-mail:lambiase@vaxsa.csied.unisa.it},
G. Scarpetta \\
{\em Dipartimento di Scienze Fisiche ``E.R. Caianiello''} \\
{\em Universit\`a di Salerno, 84081 Baronissi (SA), Italy.} \\
{\em  Istituto Nazionale di Fisica Nucleare, Sez. di Napoli.} \\
}
\date{}
\maketitle
\begin{abstract}
The maximal acceleration corrections to the Lamb shift of muonic hydrogen 
are calculated by using the relativistic Dirac wave functions. 
The correction for the
$2S-2P$ transition is $\sim 0.38$ meV and is higher than the accuracy of
present QED calculations and of the expected accuracy of experiments in
preparation. 
\end{abstract}
\thispagestyle{empty}

\vspace{20.mm}
\vspace{5.mm}
\vfill
\end{titlepage}

\section{Introduction}
\setcounter{equation}{0}

This paper presents the calculation of maximal acceleration (MA)
corrections to the
Lamb shift of muonic atoms $p^+\mu^-$, according to the model of
Caianiello and collaborators \cite{ANI}, \cite{GAS}. 
The view frequently held \cite{TUT}, \cite{BRA} 
that the proper acceleration of a particle has an upper limit finds in this
model a geometrical interpretation expressed by the line element
\begin{equation}\label{eq1}
d\tilde{s}^2=\tilde{g}_{\mu\nu}dx^{\mu}dx^{\nu}=
\left(1-\frac{|\ddot{x}|^2}{{\cal A}_m^2}\right)ds^2\equiv 
\sigma^2 (x) ds^2\,{,}
\end{equation}
experienced by the accelerating particle along its worldline. In
(\ref{eq1}) 
${\cal A}_m\equiv 2mc^3/\hbar$ is the proper MA of the particle
of mass $m$, $\ddot{x}^{\mu}=d^2x^{\mu}/ds^2$ its acceleration  and 
$ds^2=g_{\mu\nu}dx^{\mu}dx^{\nu}$ is the metric due to a background 
gravitational field. In the absence of gravity, $g_{\mu\nu}$ is replaced by the
Minkowski metric tensor $\eta_{\mu\nu}$. Results similar to (\ref{eq1})
have also been
obtained in the context of Weyl space \cite{WEY}. 

Eq. (\ref{eq1}) has several implications for relativistic kinematics
\cite{SCA}, the energy spectrum of a uniformly accelerated particle 
\cite{CGS}, the periodic structure as a function of momentum in 
neutrino oscillations \cite{CGS},
the Schwarzschild horizon \cite{TTA},  the expansion
of the very early universe \cite{SPE}, the classical electrodynamics of a 
particle \cite{ELE} and the mass of the 
Higgs boson  \cite{BOS}.
It also makes the metric observer-dependent, as conjectured by Gibbons
and Hawking \cite{GIB}, and leads in a natural way to hadron 
confinement \cite{PRE}.

The extreme large value that ${\cal A}_m$ takes
for all known particles makes a direct
test of Eq. (\ref{eq1}) very difficult. 
Nonetheless a realistic test  
has also  been suggested \cite{INI}.

Using the same model, we have recently calculated \cite{LBS} in a 
non--relativistic approximation, the MA corrections to the Lamb shift of 
hydrogenic atoms and found them compatible with experimental results. In
particular, the agreement between MA corrections and experiment is very good 
for the $2S-2P$ Lamb shift in hydrogen $(\sim 7$ kHz) and comparable 
with the agreement of experiments with standard QED with and without two--loop
corrections. The agreement also is good for the $(1/4)L_{1S}-(5/4)L_{2S}
+L_{4S}$ Lamb shift in $H$ and comparable, in some instances, with that 
between experiment and QED $(\sim 30$ kHz). The agreement remains good, in this
instance, for $D$ too. For the $L_{1S}$ case in $D$, the MA theory is worse 
($\sim -270$ kHz) than the standard one in reproducing the experimental
data when the two--loop corrections are included, but better than QED alone
when these are excluded. Finally, the MA corrections improve the agreement
between theory and experiment by $\sim 50\%$ for the $2S-2P$ shift in 
$He^+$.

In this work we extend the calculation of the MA corrections to muonic
hydrogen atoms for essentially two reasons. First, the levels of muonic
hydrogen are very sensitive to QED, recoil and proton--size effects and may 
lead to a more precise determination of the proton radius. An accurate 
measurement of the proton radius would affect all QED tests based
on the hydrogen atom and corresponding comparisons with the MA
corrections. Second, MA effects are larger in muonic hydrogen
because the muon in the ground state is much closer to the proton, hence its 
acceleration is higher. Unlike Ref. \cite{LBS}, the present calculations are
fully relativistic. Section 2 contains the Dirac Hamiltonian, its 
eigenfunctions and the MA perturbations. The Lamb shifts
are calculated in Section 3 and the conclusions are given in Section 4.

\section{The Dirac Hamiltonian}
\setcounter{equation}{0}

The MA corrections due to the metric (\ref{eq1}) appear directly in the
Dirac equation for the muon that must now be written in covariant form 
\cite{PAR} and referred to a local Minkowski 
frame by means of the vierbein field
$e_{\mu}^{\,\,\,\, a}(x)$. As in Ref. \cite{LBS} and \cite{PAR}
one finds $e_{\mu}^{\,\,\,\, a}
=\sigma(x)\delta_{\mu}^{\,\,\,\, a}$, where Latin indices refer to the locally
inertial frame and Greek indices to a generic non--inertial frame.
The covariant matrices $\gamma^{\mu}(x)$ satisfy the anticommutation 
relations ~$\{\gamma^{\mu}(x), \gamma^{\nu}(x)\}$
$=2\tilde g^{\mu\nu}(x)$, while the  covariant
derivative ${\cal D}_{\mu}\equiv \partial_{\mu}+\omega_{\mu}$ 
contains the total connection $\omega_{\mu}=
\frac{1}{2}\sigma^{ab}\omega_{\mu ab}$, where
$\sigma^{ab}=\frac{1}{4}\,[\gamma^a,\gamma^b]$, 
$\omega_{\mu\,\,\,\,b}^{\,\,\,\,a}=(\Gamma_{\mu\nu}^{\lambda}\,
e_{\lambda}^{\,\,\,\,a}-\partial_{\mu}e_{\nu}^{\,\,\,\,a})
e^{\nu}_{\,\,\,\,b}$
and  $\Gamma_{\mu\nu}^{\lambda}$
represent the usual Christoffel symbols. For conformally flat metrics
$\omega_{\mu}$ takes the form
$\omega_{\mu}=(1/\sigma)\sigma^{ab}\eta_{a\mu}\sigma_{,b}$.
By using the transformations
$\gamma^{\mu}(x)=e^{\mu}_{\,\,\,\,a}(x)\gamma^a$
so that $\gamma^{\mu}(x)=\sigma^{-1} (x)\gamma^{\mu}$,
where $\gamma ^{\mu}$ are the usual constant Dirac matrices,
the Dirac equation can be written in the form
\begin{equation}\label{eq2.1}
\left[ i\hbar\gamma^{\mu}\left(\partial_{\mu}+i\frac{e}{\hbar c}A_{\mu}\right)
+i\frac{3\hbar}{2}\gamma^{\mu}(\ln\sigma)_{,\mu}
-mc\sigma (x)\right]\psi(x)=0\,{.}
\end{equation}
>From (\ref{eq2.1}) one obtains the Hamiltonian
\begin{equation}\label{eq2.2}
H= - i\hbar c\vec{\alpha}\cdot \vec{\nabla} + e \gamma^0 \gamma^{\mu}A_{\mu}(x) 
- i\frac{3\hbar c}{2} \gamma^0 \gamma^{\mu}(\ln\sigma )_{,\mu} +
mc^2\sigma(x)\gamma^0\,{,}
\end{equation}
which is in general non--Hermitian \cite{PAR}. If $\sigma$
varies slowly in time, or is time-independent, as in the present case,
the term $(\ln\sigma )_{,0}$ can be neglected and Hermiticity is recovered.

The Lamb shift corrections are calculated by means of 
relativistic wave functions \cite{ITZ}. 
For the electric field $E(r)=kZe/r^2 (k=1/4\pi\epsilon_{0})$, the 
conformal factor becomes
$ \sigma(r)=(1-\left(\frac{r_0}{r}\right)^4)^{1/2}$,
where 
$ r_0\equiv (kZe^2/m{\cal A}_m)^{1/2}\sim 1.59\cdot 
10^{-15}\mbox{m} $
and $r>r_0$. The calculation of $\ddot{x}^{\mu}$ is performed classically.
Neglecting contributions of the order
$O({\cal A}_m^{-4})$ one gets 
$ \sigma (r)\sim 1-(1/2)(r_0/r)^4 $.
This expansion requires that in the following only those values of $r$
be chosen that are above a cut--off $\Lambda$, such that for
$r>\Lambda >r_0$ the validity of the expansion is preserved. The actual
value of $\Lambda$ is determined by the maximum probability distance of the 
muon from the proton. Thus $\Lambda\sim a_0$, where 
$a_0\equiv \hbar/m_{\mu}c\alpha$ is the Bohr radius of the muon.
The length $r_0$ has no fundamental significance in QED and depends in
general on the details of the acceleration mechanism. It is only the distance
at which the muon would attain, classically, the acceleration ${\cal A}_m$
irrespective of the probability of getting there.

By using the expansion for $\sigma (r)$ in (\ref{eq2.2}) one finds that
all MA effects are contained in the perturbative terms
\begin{equation}\label{eq8}
H_{r_0}=-\frac{mc^2}{2}\left(\frac{r_0}{r}\right)^4\beta+
i\frac{3\hbar c}{4}r_0^4\vec{\alpha}\cdot\vec{\nabla} \frac{1}{r^4}\equiv 
{\cal H}+{\cal H}^{\prime}\,{.}
\end{equation}

The corrections to the energy levels $2S$ and $2P$ are calculated by using the 
eigenfunctions of the Dirac Hamiltonian
\begin {eqnarray}\label{eq2.4}
|\psi^{(0)}> = 
	\left( \begin {array}{cc}
	g_{n_r k }(r) \chi ^{ \mu }_k \\
	if_{n_r k }(r) \chi ^{\mu }_{-k}
	\end {array} \right)\,{,}
\end {eqnarray}
where $\chi^{\mu}$ are the spin functions and
$g_{n_r k}(r)$ and $f_{n_r k}(r)$ are the radial wave functions 
\begin{equation}\label{eq2.5}
g_{n_r k}(r)=B_ke^{-\rho/2}\rho^{\gamma -1}\left[
\left(k-\frac{Z\alpha}{\lambda\lambda_c}\right){_1}{F_1}(-n_r;2\gamma +1;\rho)
+ \right.
\end{equation}
$$
\left.
+n_r{_1}{F_1}(-n_r+1;2\gamma +1;\rho)\right]\,{,}
$$
\begin{equation}\label{eq2.6}
f_{n_r k}(r)=C_ke^{-\rho/2}\rho^{\gamma -1}\left[
\left(k-\frac{Z\alpha}{\lambda\lambda_c}\right){_1}{F_1}(-n_r;2\gamma +1;\rho)
- \right.
\end{equation}
$$
\left. -n_r{_1}{F_1}(-n_r+1;2\gamma +1;\rho)\right]\,{.}
$$
${_1}{F_1}(a,b,x)$ are the confluent hypergeometric functions.
The constants in Eqs. (\ref{eq2.5}) and (\ref{eq2.6}) are defined by
\begin{equation}\label{eq2.7}
B_k\equiv A(n_r, k)\frac{(\lambda_{c}^{-1}+W)^{1/2}}{k - 
Z \alpha /\lambda\lambda_{c}},\quad 
C_k \equiv -A(n_r, k)\frac{(\lambda_{c}^{-1}-W)^{1/2}}
{k - Z \alpha /\lambda \lambda_{c}}\,{,}
\end{equation}
\begin{equation}\label{eq2.8}
A(n_r, k)\equiv \frac {2^{1/2}\lambda^{3/2}}{\Gamma (2\gamma +1)}
\left[\lambda_{c}\left(\frac{Z\alpha}{\lambda\lambda_c}
-k\right)\frac{\Gamma (2\gamma+n_r+1)}
{(\alpha /\lambda \lambda_{c})(n_r)!}\right]^{1/2}\,{.}
\end{equation}
The quantum numbers $n_r, k$ and $\gamma$ are given by
\begin{equation}\label{eq2.9}
n_r=n-\vert k\vert, \quad \gamma =\sqrt{k^2-(Z\alpha)^2}, \quad 
\alpha \sim 1/137\,{,}
\end{equation}
where $k$ is related the angular quantum number $l$
(for instance, $k=-1$ for the states $S$ and $k=1$ for the states
$P$).
$W$ is defined in terms of the energy $E_{nlj}$ 
\begin{equation}\label{eq2.10}
W=\frac{E_{nlj}}{\hbar c}=\frac{mc^2}{\hbar c}
\left[1+\frac{(Z\alpha)^2}{[n-(j+1/2)+(k^2-(Z\alpha)^2]^2}\right]^{-1/2}\,{,}
\end{equation}
where $n=1,2,3,\ldots, j=1/2, 3/2, ...\leq n, \quad 0\leq l\leq n-1$.
Finally, 
\begin{equation}\label{eq2.11}
\lambda_c=\frac{\hbar}{mc}, \quad
\lambda\equiv (\lambda_{c}^{-2}-W^{2})^{1/2}, \quad
\rho\equiv 2\lambda r\,{.}
\end{equation} 
The perturbation due to ${\cal H}^{\prime}$ vanishes, while for 
${\cal H}$ one finds
\begin {equation}\label {eq2.12}
\Delta E =-\frac {mc^{2} r_{0}^{4}}{2}
	\int^{\infty }_{\Lambda }\frac {1}{r^{2}}
	[g_{n_r k}(r)^{2}-f_{n_r k}(r)^{2}]dr\,{.}
\end {equation}

\section {$p^{+}\mu^-$ Lamb Shifts of the States $2S_{1/2}, 2P_{1/2}$ }
\setcounter{equation}{0}

The contribution to the Lamb shift $2S-2P$ is calculated by using 
Eq. (\ref{eq2.12}). 

For $ 2S_{1/2} $ states, one has (Z=1)
$n=2, n_r=1, k=-1,$ and from Eqs. (\ref{eq2.5})
and (\ref{eq2.6}) one gets
\begin{equation}\label{eq3.1}
g_{1,-1}(r) = B_{-1}e^{-\rho/2} \rho^{\gamma -1}\left[\left(1+\frac{\alpha}
{\lambda\lambda_c}\right)\frac{\rho}{2\gamma +1}-\frac{\alpha}
{\lambda\lambda_c}\right]\,{,}
\end{equation}
\begin{equation}\label{eq3.2}
f_{1, -1}(r)=C_{-1}e^{-\rho/2} \rho^{\gamma -1}\left[\left(1+\frac{\alpha}
{\lambda\lambda_c}\right)\frac{\rho}{2\gamma +1}-\frac{\alpha}
{\lambda\lambda_c}-2 \right]\,{,}
\end{equation}
where the identities \cite{STE}
\begin{equation}\label{eq3.3}
{_1}{F_1}(-1, b; x)=1-\frac{x}{b}, \quad {_1}{F_1}(0, b; x)=1
\end{equation}
have been used.
Inserting Eq. (\ref{eq3.1}) and (\ref{eq3.2}) into Eq. (\ref{eq2.12})
one obtains the correction to $2S_{1/2}$
\begin{equation}\label{eq3.4}
\Delta E(2S_{1/2})= a_{-1} I(0) + b_{-1}I(1) + c_{-1}I(2)\,{.}           
\end{equation}
The coefficients $a_{-1}, b_{-1}, c_{-1}$ and the integral function $I(q)$
are defined below. Similarly, for the state $2P_{1/2}$ one has
$n=2, n_r=1, k=1$, and
\begin{equation}\label{eq3.5}
g_{1,1}(r) = B_1e^{-\rho/2} \rho^{\gamma -1}\left[\left(-1+\frac{\alpha}
{\lambda\lambda_c}\right)\frac{\rho}{2\gamma +1}-\frac{\alpha}
{\lambda\lambda_c}+2 \right]\,{,}
\end{equation}
\begin{equation}\label{eq3.6}
f_{1, 1}(r)=C_1e^{-\rho/2} \rho^{\gamma -1}\left[\left(-1+\frac{\alpha}
{\lambda\lambda_c}\right)\frac{\rho}{2\gamma +1}-\frac{\alpha}
{\lambda\lambda_c} \right]\,{.}
\end{equation}
Therefore, the correction to the level $2P_{1/2}$ is
\begin{equation}\label{eq3.7}
\Delta E(2P_{1/2})= a_1 I(0) + b_1I(1) + c_1I(2)\,{.}           
\end{equation}
The integral function $I(q)$ is defined as
\begin{equation}\label{eq3.8}
I(q)=-mc^{2}r_{0}^{4} \lambda 
\int_{2\lambda\Lambda}^{\infty}d\rho e^{-\rho}\rho^{2\gamma -2-q}\,{,}
\end{equation}
while the constant coefficients are 
\begin{equation}\label{eq3.9}
a_{-1}=\frac{B_{-1}^2-C_{-1}^2}{(2\gamma +1)^2}\left(1+
\frac{\alpha}{\lambda\lambda_c}\right)^2\,{,}
\end{equation}
\begin{equation}\label{eq3.10}
b_{-1}=-\frac{2}{2\gamma +1}\left(1+\frac{\alpha}{\lambda\lambda_c}\right)
\left[(B_{-1}^2-C_{-1}^2)\frac{\alpha}{\lambda\lambda_c}-2C_{-1}^2\right]
\,{,}
\end{equation}
\begin{equation}\label{eq3.11}
c_{-1}=B_{-1}^2\left(\frac{\alpha}{\lambda\lambda_c}\right)^2-
C_{-1}^2\left(\frac{\alpha}{\lambda\lambda_c}+2\right)^2\,{.}
\end{equation}
\begin{equation}\label{eq3.9a}
a_{1}=\frac{B_{1}^2-C_{1}^2}{(2\gamma +1)^2}\left(-1+
\frac{\alpha}{\lambda\lambda_c}\right)^2\,{,}
\end{equation}
\begin{equation}\label{eq3.10a}
b_{1}=-\frac{2}{2\gamma +1}\left(-1+\frac{\alpha}{\lambda\lambda_c}\right)
\left[(B_{1}^2-C_{1}^2)\frac{\alpha}{\lambda\lambda_c}-2B_{1}^2\right]
\,{,}
\end{equation}
\begin{equation}\label{eq3.11a}
c_{1}=B_{1}^2\left(\frac{\alpha}{\lambda\lambda_c}-2 \right)^2-
C_{1}^2\left(\frac{\alpha}{\lambda\lambda_c}\right)^2\,{.}
\end{equation}
The integral $I(q)$ depends strongly on the cut--off $\Lambda$. 
For $\Lambda \sim a_0$, a numerical evaluation
of corrections (\ref{eq3.4}) and (\ref{eq3.7}) yields
\begin{equation}\label{eq3.12}
\Delta E(2S_{1/2})\sim - 2.06\cdot 10^5 \mbox{MHz}\,{,}
\end{equation}  
\begin{equation}\label{eq3.13}
\Delta E(2P_{1/2})\sim -2.99\cdot 10^5 \mbox{MHz}\,{,}
\end{equation}  
so that the Lamb shift correction for the muonic hydrogen atom is
\begin{equation}\label{eq3.14}
\Delta E_L=\Delta E(2S_{1/2})-\Delta E(2P_{1/2}) \sim 9.3\cdot 10^4 \mbox{MHz}
\sim 0.39 \mbox{meV}\,{.}
\end{equation}
It is interesting to note that repeating the same calculation for 
$(p^+e^-)$ hydrogen atoms, one finds
\begin{equation}\label{eq3.15}
\Delta E_L(p^+e^-)\sim 11.37\mbox{kHz}\,{,}
\end{equation}
in excellent agreement with the result $+10.45$ kHz
calculated in the non--relativistic approximation \cite{LBS}.

\section{Summary}

The results of interest in the present calculation are Eqs. (\ref{eq3.4})
and (\ref{eq3.7}). When $\Lambda\sim a_0\sim 2.6\cdot 10^{-13}$cm, the muon
Bohr radius, the $2S-2P$ Lamb shift is given by Eq. (\ref{eq3.14}). The validity
of the calculation is supported by the value (\ref{eq3.15}) obtained for the
$H$-atom, which agrees well with the result $\Delta E_L(p^+e^-)\sim
+10.45$ kHz previously calculated using a non--relativistic approximation.
The result (\ref{eq3.14}) is of opposite sign and much smaller than the 
Lamb shift from all sources $E_L=202.070(108)$ meV 
recently calculated by Pachuki \cite{PCK}, but much
higher than the estimated $0.01$ meV precision level (three--loop vacuum
polarization) of his calculation. In fact it ranks higher than all 
corrections reported in \cite{PCK} with the exception of vacuum polarization
to leading order ($205.006$ meV), two--loop vacuum polarization (1.508 meV)
and muon self--energy and vacuum polarization  (-0.668 meV). 
Measurements at the expected level 
of accuracy \cite{PCK} may provide direct evidence for the MA corrections
calculated in the present work.

\bigskip
\bigskip

\centerline{\bf Acknowledgments}

Research supported by MURST fund $40\%$ and $60\%$, DPR 382/80, the Natural
Sciences and Engineering Research Council of Canada and NATO Collaborative 
Research Grant No. 970150.
G.P. gladly acknowledges the continued research support 
of Dr. K. Denford, Dean of Science and Dr. L. Symes Vice President 
Research, University of Regina.
G.L. wishes to thank  Dr. K. Denford for his kind hospitality 
during a stay at the University of Regina.

\end{document}